\documentclass[aps,prb,twocolumn,superscriptaddress,showpacs]{revtex4-1}
\usepackage{graphicx} 	
\usepackage{graphicx,epstopdf,color}
\usepackage{amsfonts}
\usepackage{amsmath,amssymb,mathrsfs}
\usepackage{bm}
\usepackage{pst-grad}

\usepackage{dcolumn}

\begin{document}

\title{Magneto-elastically coupled structural, magnetic and superconducting
order parameters in BaFe$_{2}$(As$_{1-x}$P$_{x}$)$_{2}$}

\author{H-.H. Kuo}\altaffiliation{These authors contributed equally to this work. Correspondence addressed to analytis@slac.stanford.edu}\affiliation{Stanford Institute for Materials and Energy Sciences, SLAC National Accelerator Laboratory,\\ 2575 Sand Hill Road, Menlo Park, CA 94025, USA} \affiliation{Geballe Laboratory for Advanced Materials and Department of Materials Science and Engineering, Stanford University, USA}

\author{James G. Analytis}\altaffiliation{These authors contributed equally to this work. Correspondence addressed to analytis@slac.stanford.edu}\affiliation{Stanford Institute for Materials and Energy Sciences, SLAC National Accelerator Laboratory,\\ 2575 Sand Hill Road, Menlo Park, CA 94025, USA}\affiliation{Geballe Laboratory for Advanced Materials and Department of Applied
Physics, Stanford University, USA}

\author{J.-H. Chu}\affiliation{Stanford Institute for Materials and Energy Sciences, SLAC National Accelerator Laboratory,\\ 2575 Sand Hill Road, Menlo Park, CA 94025, USA} \affiliation{Geballe Laboratory for Advanced Materials and Department of Applied
Physics, Stanford University, USA}

\author{R. M. Fernandes}\affiliation{Department of Physics, Columbia University, New York, New York 10027,
USA}\affiliation{Theoretical Division, Los Alamos National Laboratory, Los Alamos,
NM, 87545, USA}

\author{J. Schmalian}\affiliation{Institute for Theory of Condensed Matter Physics and Center for Functional Nanostructutes, Karlsruhe Institute of Technology, Karlsruhe, 76131, Germany}

\author{I. R. Fisher}\affiliation{Stanford Institute for Materials and Energy Sciences, SLAC National Accelerator Laboratory,\\ 2575 Sand Hill Road, Menlo Park, CA 94025, USA} \affiliation{Geballe Laboratory for Advanced Materials and Department of Applied
Physics, Stanford University, USA}

\begin{abstract}
We measure the transport properties of mechanically strained single
crystals of BaFe$_{2}$(As$_{1-x}$P$_{x}$)$_{2}$ over a wide range of $x$. The N\'eel transition is extremely sensitive to stress and this sensitivity increases as optimal doping is approached, even though the transition itself is strongly suppressed. Furthermore,
we observe significant changes in the superconducting transition temperature
with applied strain, which mirror changes in the composition $x$.
These experiments are a direct illustration of the intimate coupling
between different degrees of freedom in iron-based superconductors, revealing the importance of magneto-elastic coupling to the magnetic and superconducting transition temperatures.
\end{abstract}


\maketitle

\section{Introduction}

Materials that exhibit unconventional superconductivity are almost
always in the proximity of alternative, often magnetic ground states.
Each ground state is characterized by a broken symmetry and an associated
order parameter which %
acquires a finite value at the critical temperature, indicating the
transition has occurred. The iron based superconductors fall within
a broad family of correlated electron materials which are related
to anti-ferromagnetism, joining the cuprate, heavy fermion and layered
organic superconductors. The relationship of the magnetic, structural
and other (sometimes unknown) order parameters, and particularly how
these conspire to give rise to high superconducting critical temperatures
$T_{c}$, is one of the most important experimental challenges in
understanding the mechanism behind high-temperature superconductivity.

In the present work, we reveal the intimate relationship between different
broken symmetry ground states in the BaFe$_{2}$As$_{2}$\, superconducting
family of iron-pnictides. When left chemically unmolested, these materials
are characterized by high temperature phase that is tetragonal (Tet)
and paramagnetic, transitioning at $\sim138K$ to an orthorhombic
(Ort), collinear antiferromagnet (AFM)\cite{rotter_different_2010}.
In this case, the structural transition breaks %
tetragonal symmetry ($C_{4}\rightarrow C_{2}$), %
and the shear strain $u_{xy}\equiv\partial_{y}u_{x}+\partial_{x}u_{y}$
plays the role of the order parameter. The magnetic order breaks both spin-rotational and tetragonal symmetry, characterized by an order
parameter corresponding to the 
staggered sublattice magnetization $\mathbf{M}_{i}$ where $i=1,2$
refers to the magnetization of each sublattice\cite{barzykin_role_2009,cano_interplay_2010,fernandes_effects_2010}.
When BaFe$_{2}$As$_{2}$\, is electron, hole or isovalently `doped'
\cite{rotter_different_2010,chu_determination_2009,kasahara_evolution_2010,thaler_physical_2010},
these transitions are suppressed and superconductivity (SC) emerges,
with optimal $T_{c}$ appearing when magnetism
is completely absent, indicating that AFM and SC order parameters
compete.  While for electron doped materials the structural ($T_{S}$) and
magnetic ($T_{N}$) transitions separate\cite{rotundu_first-_2011,chu_determination_2009,kuo_possible_2011}
with $T_{S}>T_{N}$, in the present isovalently substituted BaFe$_{2}$(As$_{1-x}$P$_{x}$)$_{2}$\,
materials, no such splitting is observed at any composition. 

Several theoretical descriptions have attempted to explain the %
coupling between the structural and magnetic transition, based on
pure ferroelastic phenomenology or inclusion of a nematic order parameter
\cite{barzykin_role_2009,cano_interplay_2010,Fernandes12}. 
Despite the different approaches of each of the works, all of them
highlight the importance of the magneto-elastic coupling, which can
cause the two transitions to split or to occur simultaneously. Here,
 we investigate the role of the magneto-elastic coupling by studying the thermodynamic response of the material BaFe$_{2}$(As$_{1-x}$P$_{x}$)$_{2}$ to a small mechanical strain applied along the tetragonal {[}110]$_{T}$ direction, or equivalently the orthorhombic $b$ axis when $T<T_N$. We find that
a small shear stress $\sigma$ can significantly alter the N\'eel transition
$T_{N}$ and superconducting $T_{c}$, in a manner akin to changing
$x$. Surprisingly, even though the magnetism is almost completely
suppressed at optimal doping, the effect on the magnetic transition
grows, suggesting that magneto-elastic fluctuations substantially
increase near optimal $T_{c}$.

\section{Experimental Methods}

\begin{figure*}
\includegraphics[width=13cm]{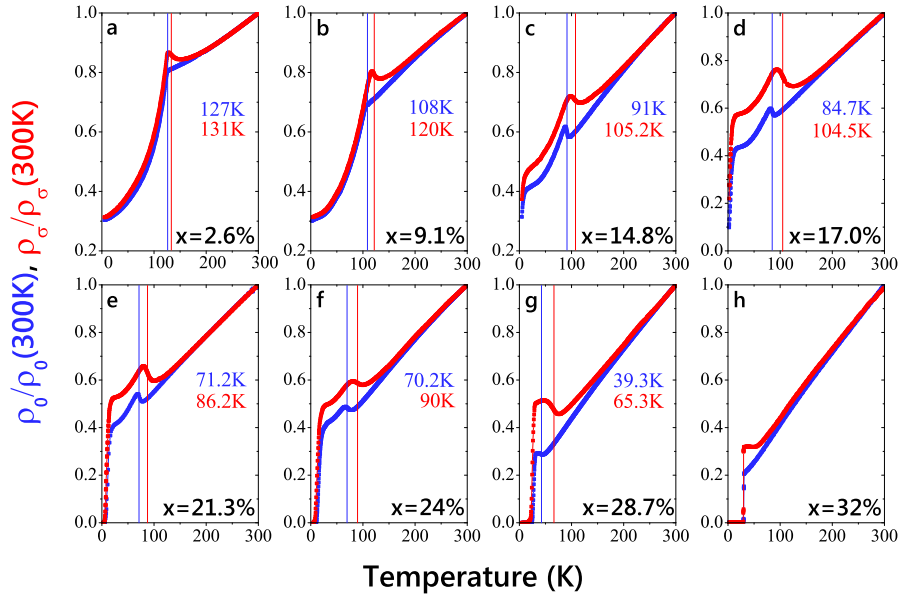} 

\caption{Temperature dependence of the normalised resistivity, $\rho/\rho(300K)$,
for unstressed (blue), and uni-axially stressed along [110]$_T$ (red) BaFe$_{2}$(As$_{1-x}$P$_{x}$)$_{2}$\ crystals
with $x=2.6\%$ to $x=32\%$. $T_{N}$ determined from $d\rho/dT$
are labeled and denoted by vertical lines with corresponding colors (see Appendix \ref{TNderiv} for details).}

\label{uniax_RT} 
\end{figure*}

The growth of single crystals of BaFe$_{2}$(As$_{1-x}$P$_{x}$)$_{2}$\, is described elsewhere \cite{analytis_enhanced_2010}. It is worth noting however that the quality of the crystals can be improved by annealing within the growth for a week at 900$^\circ$C.
Samples were mechanically strained along either the {[}100]$_{T}$
or {[}110]$_{T}$ direction using a custom built mechanical device
described in Ref. \cite{chu_-plane_2010}. A cantilever was pressed
against the sample by adjusting a turnable screw, applying $<10MPa$.
Each sample was cut to have similar dimensions (300$\times$200$\times$80$\mu m^{3}$
- the largest dimension of the batch with the smallest samples). To
reduce the errors associated with differences in pressure, the experiments
were repeated on 3 samples from each batch. This gives us confidence
that we are able to apply a similar stress for all samples and hence
that the changes we detect between samples from different batches
(different P content $x$) are in fact systematic. This study is distinct
from our earlier investigations of the transport anisotropy, which
is a non-equilibrium property, whereas we presently focus to the effect
of mechanical strain on the temperature of the phase transitions themselves.

\section{Results}

Figure \ref{uniax_RT} (a)-(h) illustrates the main experimental data.
In each panel we show the normalised resistivity vs temperature for
an unstressed (blue) and stressed (red) crystal at a given doping,
where the stress has been applied along the {[}110]$_{T}$. The vertical
lines denote the assigned N\'eel transitions for each curve which have
been determined by the minimum in the resistivity derivative (Appendix \ref{TNderiv}).
Note that in contrast to electron doped materials where two anomalies
are observed in $d\rho/dT$ in the \textit{unstrained} samples, we only observe one in BaFe$_{2}$(As$_{1-x}$P$_{x}$)$_{2}$,
indicating that $T_{N}\approx T_{s}\,$for all$\,x$ \cite{chu_determination_2009}. This may suggest that the magneto-elastic coupling in these materials is
perhaps larger \cite{cano_interplay_2010,Fernandes12}, though we point out that other studies have observed split transitions in these compounds \cite{kasahara_evolution_2010}. In the presence
of mechanical strain, the structural transition will naturally broaden
to higher temperatures, but by continuity, the anomaly seen in the
data of mechanically strained samples must be associated with the
magnetic order.

In all the samples we studied, $\Delta T_{N}=T_{N}(\sigma)-T_{N}(\sigma=0)>0$, where $\sigma$ indicates
the mechanical strain field (the stress). Intriguingly, for the unstressed
optimally superconducting $x=32\%$ sample (blue curve in Figure \ref{uniax_RT}
(h)) there is no detectable magnetic transition, but after application
of stress a distinct minimum arises at $45K$, almost identical to
the minimum seen in the unstressed samples at lower doping (consider
blue curve in Figure \ref{uniax_RT} (g)). It appears that the magnetic
transition has been summoned from beneath the superconducting dome
by the application of mechanical strain. Even though the magnetic
order itself vanishes, the strong magneto-elastic coupling as well
as magnetic and elastic fluctuations remain.

\begin{figure}
\includegraphics[width=8.5cm]{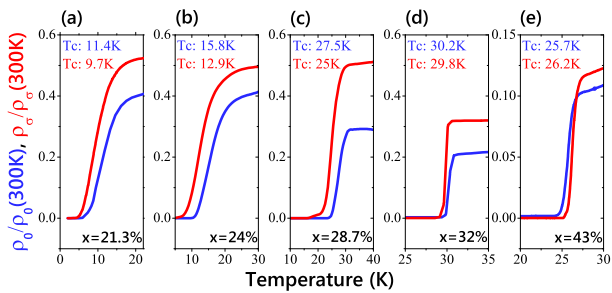} 

\caption{Expanded view of the data shown in Fig \ref{uniax_RT} around to $T_{c}$
for $x=21.3\%$ to $x=32\%$ ((a) to (d),underdoped) and $x=21.3$ $\%$
((e),overdoped). Red and blue curves correspond to the normalized resistivity
of unstressed, and uni-axially stressed (along [110]$_T$) respectively. For underdoped samples, $\Delta T_{c}>0$;
for overdoped samples, $\Delta T_{c}<0$.}

\label{uniax_Tc} 
\end{figure}

In Figure \ref{uniax_Tc} we show the same data presented in Figure
\ref{uniax_RT}, but focus around the superconducting critical temperature
at each composition. In this case $T_{c}$ is defined as the midpoint
in the superconducting transition. Even though the transition likely
gets broader with the application of strain, the superconducting $T_{c}$
nevertheless can be seen to \textit{decrease} as strain is applied
until compositions beyond optimal $x$, where we observe $T_{c}$
to increase (Figure \ref{uniax_Tc} (e) illustrates that $\Delta T_{c}=T_{c}(\sigma)-T_{c}(\sigma=0)$
switches sign beyond optimal $x$). Finally, the affect on $T_{N}$ and $T_{c}$
is proportional to the amount of applied strain, and this can be demonstrated
by applying a systematically increasing amount (Appendix \ref{TNcomp}). These data are suggestive that the
application of shear stress has a similar affect to decreasing $x$
across the phase diagram.

\begin{figure}
\includegraphics[width=8.5cm]{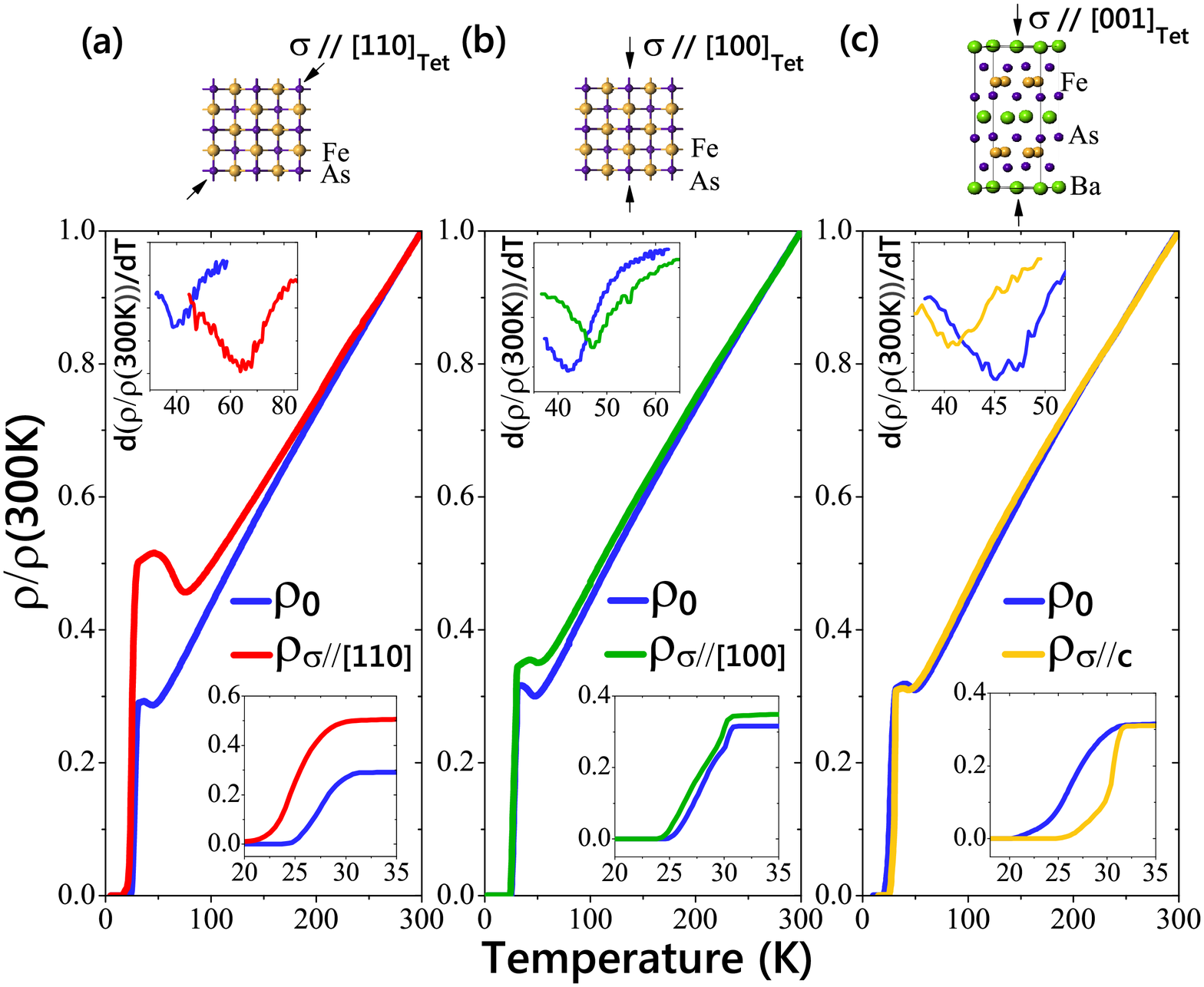} 

\caption{Temperature dependence of normalized in-plane resistivity $\rho/\rho(300K)$,
for unstressed crystals and (a) uni-axially stressed along {[}110]$_{T}$
(red), (b) uni-axially stressed along {[}100]$_{T}$ (green), and
(c) uni-axially stressed along the crystal c axis (yellow) BaFe$_{2}$(As$_{1-x}$P$_{x}$)$_{2}$
crystals with x=28.7$\%$. The upper inserts show $d\rho/dT$ vs T
of each crystal around $T_{N}$ with corresponding colors. For (a),
$\Delta T_{N}$ $\sim27K$, (b), $\Delta T_{N}$ $\sim5K$, and (c),
$\Delta T_{N}$ $\sim-5K$. The lower incepts are magnified plots
of $\rho$ vs T around $T_{c}$. For (a), $\Delta T_{c}$ $\sim-2.5K$,
(b), $\Delta T_{c}$ $\sim-0.3K$, and (c), $\Delta T_{c}$ $\sim4K$.}

\label{tet} 
\end{figure}

To ensure that this effect is indeed intrinsic to pressure along one
of the orthorhombic axes, we also apply pressure along {[}100]$_{T}$, shown for
comparison on two samples from the same batch in Figure \ref{tet}
(a) and (b). Mechanical strain in this direction results in small
changes in both magnetic and superconducting transitions. We furthermore
apply pressure in the inter-layer $c$ direction (Figure \ref{tet}
(c)) and found the opposite behavior whereby $\Delta T_{N}<0$ and
$\Delta T_{c}>0$. This suggests that the ratio $c/b$ whether directly
or indirectly, likely plays a role in the superconducting mechanism.

\section{Discussion}

\begin{figure}
\includegraphics[width=9.5cm]{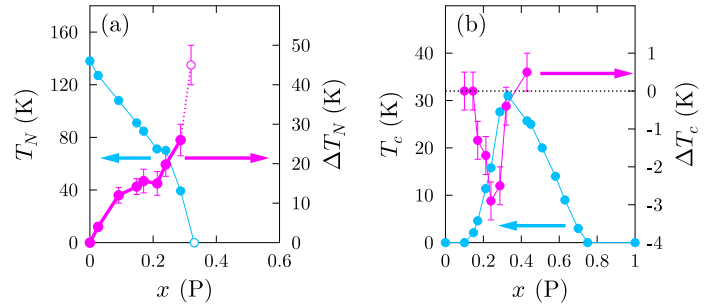} 
\caption{(a) $T_{N}(\sigma=0)$ (blue, left axis) and $\Delta T_{N}$ (pink,
right axis) vs P concentration $x$. Points at optimal doping are
distinguished as open circles because they rely on the assumption
that $T_{N}(\sigma=0)=0$. (b) $T_{c}(\sigma=0)$ (blue, left axis)
and $\Delta T_{c}$ (pink, right axis) vs P concentration $x$. Dotted
black line indicates $\Delta T_{c}=0$. In both (a) and (b) the source
of error is predominantly related to the our ability to accurately
determine minima in the resistivity derivative about the N\'eel\, and
superconducting transitions. The size of the effect we see on each
of these transitions is confirmed on multiple samples.}

\label{DELTATN} 
\end{figure}

In Figure \ref{DELTATN} (a) we plot the change in the N\'eel temperature
$\Delta T_{N}$ as a function of doping for nominally the same %
 stress at each doping. Surprisingly, even though the magnetic and
the structural transitions are suppressed as a function of doping,
the amount that $T_{N}$ can be changed by stress monotonically \textit{increases}
with P content, within our error bars. This is in contrast
to the value of the resistivity anisotropy itself, which is highly
non-monotonic with doping (Appendix \ref{resaniso}). If we also include data
of $T_{N}(\sigma)$ at $x=32\%$, $\Delta T_{N}$ appears to have
the largest response at optimal doping, as shown in Figure \ref{DELTATN}
(a). This large enhancement of $\Delta T_{N}$ implies an enhanced
susceptibility of the AFM ground state to shear stress.

In Figure \ref{DELTATN} (b), we illustrate the equivalent plot of
changes in the superconducting critical temperature $\Delta T_{c}$
as a function of doping for nominally the same strain. $\Delta T_{c}$,
unlike $\Delta T_{N}$, is not monotonic with $x$. Comparing the dependence with the evolution of the {\it unstressed} superconducting transition with $x$, it appears that the magnitude of $\Delta T_{c}$ is largest
where the dome is steepest and small otherwise; mathematically $\Delta T_{c}\propto-dT_{c}/dx$. Indeed,
considering Figure \ref{DELTATN} (a) a very similar equation likely
applies to the magnetic transition, so that $\Delta T_{N}\propto-dT_{N}/dx$.
As $T_{N}$ is increased, there are likely fewer electrons available
to participate in superconductivity\cite{Vorontsov09,Fernandes_Schmalian},
 and so the fact that $T_{c}$ decreases with applied stress in the
underdoped region is not surprising, since $T_{N}$ increases. However,
we note than an unstrained sample with a given $T_{N}$ has always
a $T_{c}$ that is lower than a strained sample with the same $T_{N}$,
i.e. $T_{c}(T_{N},0)<T_{c}(T_{N},\sigma)$. Therefore, there is an
intrinsic effect of stress on $T_{c}$, beyond the indirect effect
due to the competition between AFM and SC. 

We employ a Ginzburg-Landau (GL) approach to obtain further insight into
our observations. The GL model has been applied to the ferro-pnictides
by several authors already to describe the coupling between structure
and magnetism\cite{barzykin_role_2009,mazin_key_2009,cano_interplay_2010,fernandes_effects_2010,paul_magnetoelastic_2011,kim_character_2011}. From symmetry considerations, the coupling between magnetic and elastic
degrees of freedom enters the free energy via: 
\begin{equation}
F_{ME}=g\left (\mathbf{M}_{1}\cdot\mathbf{M}_{2}\right)u_{xy}.
\label{Gme}
\end{equation}
where $g$ is the magneto-elastic coupling. In order to describe the
present experiment, we need to add also the term $-\sigma u_{xy}$,
where $\sigma$ is the applied mechanical stress. If one assumes that
the structural and magnetic transitions occur independently, the coupling
(\ref{Gme}) effectively ties them together \cite{cano_interplay_2010}.
Alternatively, it has been proposed that the structural transition
is a secondary consequence of an underlying electronic order dubbed
nematic \cite{chu_-plane_2010}. In this case, an independent order
parameter $\eta\propto\left\langle \mathbf{M}_{1}\cdot\mathbf{M}_{2}\right\rangle $
condenses and triggers the structural transition via the coupling
(\ref{Gme}). Within this approach, the elastic degrees of freedom
are not intrinsically soft and can be integrated out from the partition function
(see Supplemental Material for more details), yielding the contribution
to the free energy $\propto\frac{g\sigma}{C_{s}^{0}}\left(\mathbf{M}_{1}\cdot\mathbf{M}_{2}\right)$, where $C_{s}^{0}$ is the bare elastic shear constant
(see also \cite{canonew,kivelsonnew}). This term shows that the mechanical
stress is converted into a conjugate field $\sigma g/C_{s}^{0}$ to the
electronic order parameter $\eta$. Furthermore, it also changes the
magnetic part of the free energy, resulting in an increase of the
magnetic transition temperature $T_{N}(\sigma\ne0)>T_{N}(\sigma=0)$.

As $x$ approaches optimal compositions, we observe that the magneto-elastic
response is enhanced, which suggests that $g/C_{s}^{0}$ increases substantially
and is strongest at optimal doping. In contrast, our previous mechanical
strain studies on Ba(Fe$_{1-x}$Co$_{x}$)$_{2}$As$_{2}$\, did
not show significant changes in $T_{N}$, and changes have only been
observed for pressures $\sim$5$\times$ those used here \cite{liang_effects_2011},
though a recent neutron study in the parent compound could detect
changes in $T_{N}$ at much smaller pressures \cite{dhital_effect_2011}.
Nevertheless, as a function of Co doping the difference between $T_{N}$
and $T_{S}$ grows\cite{rotundu_first-_2011,chu_determination_2009},
which could be interpreted in terms of a decreasing $g$ \cite{cano_interplay_2010,Fernandes12}.
Furthermore, even though the lattice softens as a function of temperature,
the average value of $C_{s}^{0}$ in fact increases as a function of
Co doping \cite{fernandes_effects_2010,yoshizawa_structural_2012}.
Here, in contrast, the effect of stress on $T_{N}$ is strongly enhanced
as $x$ increases in BaFe$_{2}$(As$_{1-x}$P$_{x}$)$_{2}$, suggesting
that either $g$ becomes larger or $C_{s}^{0}$ smaller, or both. %
Another possibility is that, near optimal doping, where there is no
structural or magnetic transitions, the nematic susceptibility $\chi_{\mathrm{nem}}$
is enhanced, providing an additional contribution that enhances the
{}``conjugate field'' $\frac{g\sigma}{C_{s}^{0}}\chi_{\mathrm{nem}}\left(\mathbf{M}_{1}\cdot\mathbf{M}_{2}\right)$
and, consequently, $\Delta T_{N}$ (see Appendix \ref{GLA}).
Interestingly, experiments have indicated that magnetic fluctuations
are critical at optimally-doped BaFe$_{2}$(As$_{1-x}$P$_{x}$)$_{2}$
\cite{kasahara_evolution_2010}, which could suggest a close connection
between nematic and magnetic fluctuations in these compounds \cite{fernandes_effects_2010,paul_magnetoelastic_2011}.

Furthermore, magnetic fluctuations enhance the repulsive inter-band
pairing interaction that can lead to an unconventional superconducting
state \cite{Mazin08}. Nematic fluctuations, on the other hand, give
rise to an attractive intra-band pairing interaction, which can potentially
enhance the transition temperature of the unconventional SC phase
\cite{fernandes2012}.

Previous x-ray studies on Ba(Fe$_{1-x}$Co$_{x}$)$_{2}$As$_{2}$
showed that $u_{xy}$ is strongly suppressed below $T_{c}$ \cite{nandi_anomalous_2010},
indicating that the SC and orthorhombic phases compete. One would
then expect that by applying stress and inducing a non-zero $u_{xy}$,
$T_{c}$ would decrease. However, our observations that $T_{c}\left(\sigma\right)>T_{c}\left(0\right)$
in the overdoped region and $T_{c}(T_{N},\sigma)>T_{c}(T_{N},0)$
in the underdoped region suggest that the applied stress may actually
lead to an intrinsic increase of $T_{c}$. To understand the effect
of mechanical stress on superconductivity, we can compare to the case
of high-$T_{c}$ copper-oxide based materials \cite{meingast_superconducting_1996,locquet_doubling_1998,hardy_enhancement_2010}.
Though the effects vary between different compounds, Hardy \textit{et
al.} proposed a unified picture of the influence of uni-axial strain
(excluding YBCO), whereby changes in $T_{c}$ could be accounted for
by changes in the ratio $c/a$ \cite{hardy_enhancement_2010}. In
the present study, $c/b$ will always increase when stress is applied
along {[}110]$_{T}$, and by a smaller amount when applied along {[}100]$_{T}$,
but will decrease when applied along $c$ (see Figure \ref{tet}).
The changes we are able to invoke on the under-doped samples follow
this trend, so that qualitatively $\Delta T_{c}\propto-\Delta(c/b)$.
However, one cannot say whether it is the lattice parameters alone,
their ratio or some other systematically adjusted internal parameter
which is most important (such as the As-Fe-As bond angle). Direct
structural measurements as a function of mechanical strain are required
to answer this question.

\section{Conclusions}

In conclusion, we have found a strongly enhanced magneto-elastic response
in BaFe$_{2}$(As$_{1-x}$P$_{x}$)$_{2}$\,as $x$ approaches optimal
doping, which may be related to the superconducting mechanism itself.
We also find that mechanical strain can directly couple to the superconducting
order parameter in a manner that is similar to decreasing the P concentration
$x$. These experiments are therefore a direct illustration of the
subtle coupling between different degrees of freedom in BaFe$_{2}$(As$_{1-x}$P$_{x}$)$_{2}$. 

\section{Acknowledgments}

RMF acknowledges support of NSF Partnerships for International Research and Education (PIRE) program OISE-0968226. HHK, JGA, JHC and IRF acknowledge support of the U.S. DOE, Office of Basic Energy Sciences under contract DE-AC02-76SF00515.

\appendix

\section{Determination of the N\'eel temperature from the resistivity derivative}
\label{TNderiv}

\begin{figure*}
\includegraphics[width=14cm]{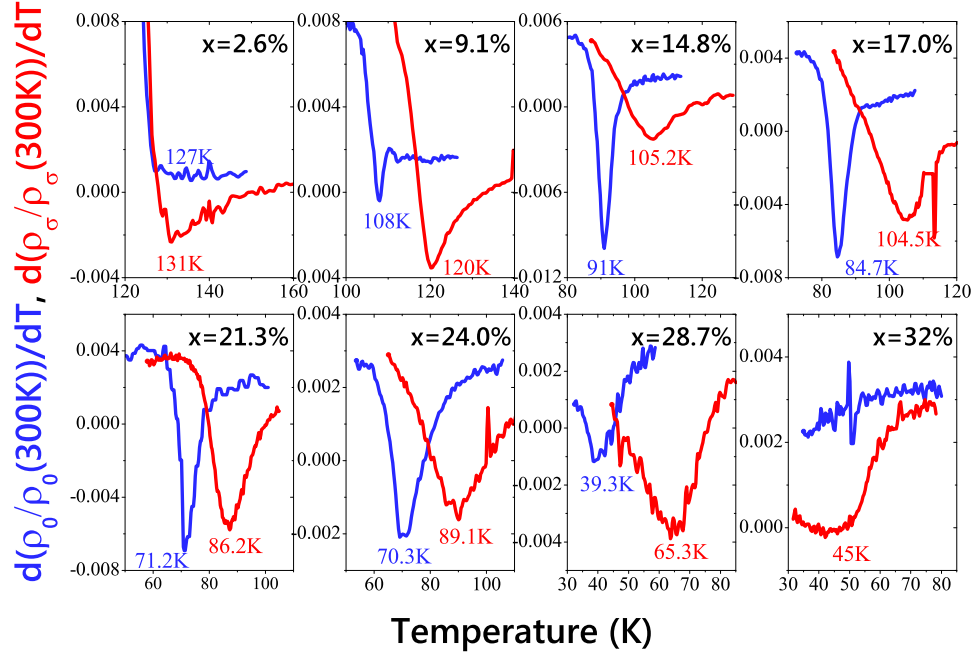} 

\caption{Derivative of normalised resistivity with respect to temperature,
${d\over dT} \rho/\rho(300K)$, for unstressed (blue), and uni-axially stressed
along [110]$_T$ (red) BaFe$_{2}$(As$_{1-x}$P$_{x}$)$_{2}$\ crystals
with $x$=2.6$\%$ to $x$=32$\%$. The minimum in $d\rho/dT$
is shown in corresponding color, and is interpreted as the N\'eel temperature $T_{N}$.}
\label{drdt}
\end{figure*}

The Fermi surface reconstruction associated with the N\'eel order at $T_{N}$ appears as a pronounced minimum in the derivative with respect to temperature. After applying strain, we observe an increase in $T_{N}$, as shown in Fig. \ref{drdt}. Blue and red represent unstressed and uni-axially stressed (along [110]$_T$) BaFe$_{2}$(As$_{1-x}$P$_{x}$)$_{2}$ single crystals respectively. Even though the stress broadens the transition, the increase in $T_N$ can be easily resolved.

\section{Systematic response of $T_{N}$ to pressure}
\label{TNcomp}

\begin{figure}
\includegraphics[width=7cm]{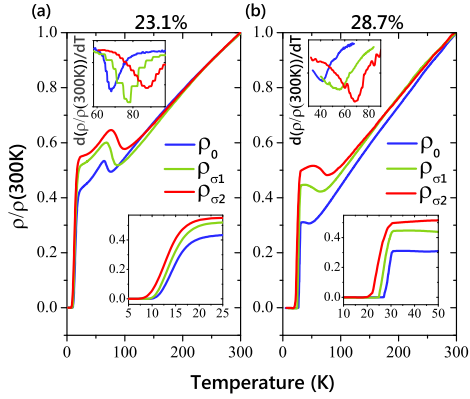} 

\caption{ Temperature dependence of normalized resistivity of BaFe$_{2}$(As$_{1-x}$P$_{x}$)$_{2}$,
$x$ = 23.1\% and $x$ = 28.7\%, at three systematically increased amounts of stress. The normalized resistivity $\rho_{i}$ with $i = \{ 0, \sigma 1, \sigma 2 \}\,$ stand for unstressed (blue), intermediately stressed
(green) and highly stressed (red) respectively. The upper inserts show
$d\rho_{i}/dT$ vs T near $T_{N}$, and the lower inserts are expanded
plots of $\rho_{i}$ near $T_{c}$.}

\label{pressureD} 
\end{figure}

Although we cannot determine the absolute value of stress applied,
we can nevertheless tune the amount of pressure applied by gradually
tightening the screw on the device. As a typical example, two samples of BaFe$_{2}$(As$_{1-x}$P$_{x}$)$_{2}$,
$x$ = 23.1\% and $x$ = 28.7\% are shown in Fig. \ref{pressureD}. Each sample was measured with a different amount
of stress applied along [110]$_T$. $|\Delta T_{N}|$
and $|\Delta T_{c}|$ increase with increasing stress.

\section{Resistivity Anisotropy}
\label{resaniso}

The in-plane resistivity anisotropy of BaFe$_{2}$(As$_{1-x}$P$_{x}$)$_{2}$ was obtained by measuring the resistivity of mechanically detwined crystals as described elsewhere \cite{kuo_possible_2011}. Clearly, the resistivity anisotropy has a highly non-monotonic dependece on doping. This is in contrast to the  
trend of the response of $T_N$ at constant stress as a function of doping, which is a monotonic increase with the concentration $x$, as shown in Figure \ref{DELTATN}.  

\begin{figure}[ht]
\includegraphics[width=7cm]{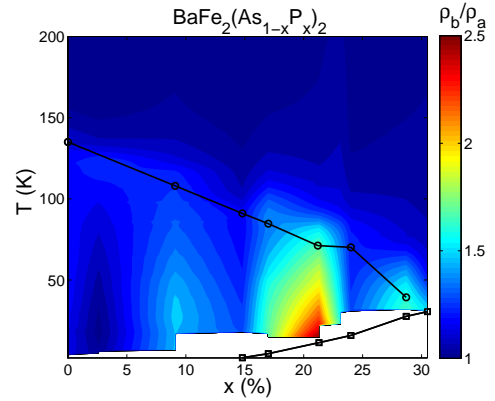}
\caption{ In-plane resistivity anisotropy $\rho_{a}/\rho_{b}$ as a function of temperature and doping for BaFe$_{2}$(As$_{1-x}$P$_{x}$)$_{2}$. The color scale has been obtained by a linear interpolation between adjacent data points. Black circles and squares indicate $T_s/T_N$ and $T_c$ respectively.}

\label{ba} 
\end{figure} 

\section{Ginzburg-Landau analysis}
\label{GLA}
To understand how the different degrees of freedom affect $T_{N}$
in mechanically stressed samples, we use a phenomenological Ginzburg-Landau
model for the magnetic, nematic, and elastic degrees of freedom. For
the magnetic part, we have:

\begin{equation}
F_{mag}=\frac{r_{0}}{2}\left(M_{1}^{2}+M_{2}^{2}\right)+\frac{u}{8}\left(M_{1}^{2}+M_{2}^{2}\right)^{2}-\frac{\lambda}{2}\left(\mathbf{M}_{1}\cdot\mathbf{M}_{2}\right)^{2}
\label{magnetic_F}
\end{equation}

where $\mathbf{M}_{1}$ and $\mathbf{M}_{2}$ are the staggered magnetization
of the two interpenetrating N\'eel sublattices. Here, we defined $r_{0}=a\left(T-T_{N,0}\right)$,
with $T_{N,0}$ denoting the mean-field magnetic transition temperature.
The coupling constants satisfy $u>0$ and $\lambda>0$, such that
in the free-energy minimum $\mathbf{M}_{1}$ and $\mathbf{M}_{2}$
are either parallel or anti-parallel, corresponding to the two possible
magnetic stripe configurations with ordering vectors $\left(\pi,0\right)$
and $\left(0,\pi\right)$ in the 1-Fe unit call Brillouin zone. It is convenient to introduce the order
parameters of these two magnetic stripe states $\boldsymbol{\Delta}_{1}$
and $\boldsymbol{\Delta}_{2}$, such that $\boldsymbol{\Delta}_{1}=\left(\mathbf{M}_{1}+\mathbf{M}_{2}\right)/2$
and $\boldsymbol{\Delta}_{2}=\left(\mathbf{M}_{1}-\mathbf{M}_{2}\right)/2$.
Notice that $M_{1}^{2}+M_{2}^{2}=2\left(\Delta_{1}^{2}+\Delta_{2}^{2}\right)$
and $\mathbf{M}_{1}\cdot\mathbf{M}_{2}=\Delta_{1}^{2}-\Delta_{2}^{2}$.

We now consider the nematic part in a phenomenological way. The nematic
order parameter $\varphi$ breaks the tetragonal symmetry of the lattice.
At high enough temperatures (compared to the structural transition
temperature), such as those for the optimally-doped compounds, we
consider the free energy expansion only up to second-order in the
nematic order parameter:

\begin{equation}
F_{nem}=\frac{1}{2}\left(\chi_{nem}^{(0)}\right)^{-1}\varphi^{2}-\kappa\varphi\left(\Delta_{1}^{2}-\Delta_{2}^{2}\right)\label{nematic_F}\end{equation}
where $\kappa$ is the coupling between nematic and magnetic degrees
of freedom, and $\chi_{nem}^{(0)}$ is the static nematic susceptibility.
For the elastic part, we consider a harmonic lattice

\begin{equation}
F_{el}=\frac{C_{s}^0}{2}u_{xy}^{2}-gu_{xy}\varphi-\sigma u_{xy}\label{elastic_F}\end{equation}
where $g$ is the magneto-elastic coupling, $C_{s}^0$ is the bare shear
modulus, and $\sigma$ is the applied stress. 

To study how $T_{N,0}$ changes as function of $\sigma$, we first
integrate out the elastic degrees of freedom from the partition function

\begin{eqnarray}
\int du_{xy}\,\mathrm{e}^{-\frac{C_{s}^0}{2}u_{xy}^{2}+u_{xy}\left(g\varphi+\sigma\right)} & \propto\nonumber \\
\exp\left[\frac{\left(g\varphi+\sigma\right)^{2}}{2C_{s}^{0}}\right]
\label{aux1}
\end{eqnarray}

Substituting in Eq. (\ref{nematic_F}), the nematic free energy becomes

\begin{eqnarray}
F_{nem} & = & \frac{\chi_{nem}^{-1}}{2}\varphi^{2}-\varphi\left[\kappa\left(\Delta_{1}^{2}-\Delta_{2}^{2}\right)+\frac{g\sigma}{C_{s}^{0}}\right]
\label{Fnem_renormalized}
\end{eqnarray}
where we defined the renormalized nematic susceptibility $\chi_{nem}^{-1}=\left(\chi_{nem}^{(0)}\right)^{-1}-\frac{g^{2}}{2C_{s}^{0}}$.
If we consider the regime where the nematic free energy can be approximated
by the quadratic expansion (\ref{nematic_F}), we can also integrate
out the nematic degrees of freedom, obtaining

\begin{eqnarray}
\int d\varphi\,\mathrm{e}^{-\frac{\chi_{nem}^{-1}}{2}\varphi^{2}+\varphi\left[\kappa\left(\Delta_{1}^{2}-\Delta_{2}^{2}\right)+\frac{g\sigma}{C_{s}^{0}}\right]} & \propto\nonumber \\
\exp\left[\frac{\left(\kappa\left(\Delta_{1}^{2}-\Delta_{2}^{2}\right)+\frac{g\sigma}{C_{s}^{0}}\right)^{2}}{2\chi_{nem}^{-1}}\right]\label{aux3}\end{eqnarray}

Substituting in Eq. (\ref{magnetic_F}), the magnetic free energy
becomes

\begin{eqnarray}
F_{mag} & = & r_{0}\left(\Delta_{1}^{2}+\Delta_{2}^{2}\right)+\frac{u}{2}\left(\Delta_{1}^{2}+\Delta_{2}^{2}\right)^{2}\nonumber \\
 &  & -\frac{1}{2}\left(\lambda+\frac{\kappa^{2}}{\tilde{\chi}_{nem}^{-1}}\right)\left(\Delta_{1}^{2}-\Delta_{2}^{2}\right)^{2}\nonumber \\
 &  & -\frac{g\kappa\sigma}{\chi_{nem}^{-1}C_{s}^{0}}\left(\Delta_{1}^{2}-\Delta_{2}^{2}\right)\label{Fmag_renormalized}\end{eqnarray}

The last term acts as a conjugate field and breaks the tetragonal
symmetry, selecting the magnetic stripe configuration corresponding
to the $\Delta_{1}$ order parameter (ordering vector $\left(\pi,0\right)$).
Since $r_{0}=a\left(T-T_{N,0}\right)$, the magnetic transition temperature
is given by

\begin{eqnarray}
T_{N} & = & T_{N,0}+\left(\frac{g\kappa\chi_{nem}}{aC_{s}^{0}}\right)\sigma\label{TN}\end{eqnarray}

Hence, the increase in $T_{N}$ is proportional to the applied strain
$\sigma$. The enhanced response at optimal doping can be due to one
(or a combination) of the following features: an intrinsic softening
of the lattice (i.e. decrease $C_{s}^{0}$), an enhancement to the magneto-elastic
coupling (i.e. increase of $g$ and/or $\kappa$), and an enhancement
of nematic fluctuations (i.e. increase of $\chi_{nem}$). A similar enhancement in $T_N$ is also expected even in the absence of a nematic order parameter, as pointed out recently by Cano and Paul \cite{canonew}.


\end{document}